\documentclass{PoS}

\def\rz{\rm I\kern-.2emR}
\def\gz{{\rm Z\kern-.48em Z}}

\def\be{\begin{equation}}
\def\ee{\end{equation}}
\def\bea{\begin{eqnarray}}
\def\eea{\end{eqnarray}}

\title{Critical discussion of Tomboulis's approach to the confinement 
problem}

\ShortTitle{Tomboulis's approach}

\author{Keiichi R. Ito\\
        Department of Mathematics and Physics\\
        Setsunan University\\
        Neyagawa, Osaka 572-8508, Japan\\
        E-mail: \email{ito@mpg.setsunan.ac.jp}}

\author{\speaker{Erhard Seiler}\\
        Max-Planck-Institut f\"ur Physik\\ (Werner-Heisenberg-Institut)\\
        F\"ohringer Ring 6\\
        D-80805 M\"unchen, Germany\\
        E-mail: \email{ehs@mppmu.mpg.de}}

\abstract{The approach to the confinement  problem proposed by E.T. 
Tomboulis is analyzed critically. We point out some problems  
and discuss the chances of producing a proof by using his strategy.}

\FullConference{8th Conference Quark Confinement and the Hadron Spectrum \\
		 September 1-6 2008\\
		 Mainz, Germany}

\begin{document}

\section{Historical remarks}
Since this talk is about a very old issue, it seems appropriate to start 
with a few historical remarks to put the problem and its proposed solution 
by Tomboulis into context.

The confinement problem in lattice Yang-Mills theory was a hot issue in 
the late 1970's and early 1980's. Center vortices were identified by 
several authors as crucial objects (\cite{yoneya, thooft, mp}). 't~Hooft 
proposed a confinement criterion inspired by these vortices; unlike the 
earlier criterion proposed by Wilson \cite{wilson} and its modification 
by Polyakov \cite{polyakov} it did not involve infinitely heavy quark 
sources but (sourceless) central electric flux in a torus.

A little later it was proven that 't Hooft's confinement criterion implies 
confinement in the sense of Wilson and Polyakov \cite{bs,ty}. 

About the same time Tomboulis \cite{tombprl} came up with a charming idea 
how to prove that lattice Yang-Mills theory based on a nonabelian 
(compact, semisimple) gauge group has a nonzero string tension in 
't~Hooft's sense at all values of the bare coupling constant: he proposed 
to link by rigorous inequalities lattice Yang-Mills theory to the 
solution of an approximate Renormalization invented earlier by Migdal and 
Kadanoff (MK RG) \cite{mk, jose}.

It was proven a little later that in 4 dimensions the MK RG drives 
lattice $SU(N)$ Yang-Mills theory, but also compact lattice QED to the 
strong coupling fixed point \cite{ito}. This signals confinement for 
these models, and is therefore misleading for the abelian model, which is 
known to have a deconfining transition \cite{guth, fs}.

This fact raised problems for Tomboulis's approach, because it was not 
clear how his inequalities would distinguish between the groups $SU(N)$ 
and $U(1)$, especially since in his short letter there were no details 
given concerning the proof of the crucial inequalities. In fact there 
even remained room for doubt as to the existence of confinement in this 
sense in the $SU(N)$ lattice models or the analogous question of mass 
generation in $2D$ $O(N)$ spin models (see for instance \cite{seiler} and 
references given there). In spite of these efforts as well as the efforts 
by others, such as K.~R.~Ito \cite{itoconf}, who tried to prove 
mathematically the correctness of the common expectations, neither 
confinement for arbitrarily weak bare coupling nor its absence could be 
established (nor could the analogous $2D$ problem be definitely settled). 
The problem remains an important open question to this day.

In 2007 Tomboulis \cite{tomb07} revived his old idea (with some 
modifications) and published a paper providing details about the 
purported proof. The following remarks, while critical of his work, 
should nevertheless not diminish his credit for having revived interest 
in this old, important but neglected and unsolved problem. A more 
detailed discussion can be found in our paper \cite{is}.
 
\section{Sketch of Tomboulis's strategy}
The goal of Tomboulis's strategy, for simplicity for the gauge group 
$SU(2)$,  is to establish the spreding of central magnetic flux on a 
torus $\Lambda$ of dimensions $L_1\times L_2\times L_3 \times L_4$:

\be
\frac{Z^{(-1)}_\Lambda}{Z_\Lambda}\ge
\exp\left[-cL_2 L_3 e^{-\alpha L_1 L_4}\right]
\quad {\rm for} \quad L_1 L_4 \gg \log (L_2 L_3)\,,
\label{ineq}
\ee
where $Z^{(-1)}_\Lambda$ has twisted boundary conditions in the (12) 
direction. (\ref{ineq})is  supposed to follow from 
\be
\frac{Z_\Lambda^{(-)}}{Z_\Lambda}\ge
\frac{Z^{(-)}_{MKT}(n)}{Z_{MKT}(n)}\,.
\label{MKTineq}
\ee
Here $Z_{MKT}(n)$, $Z^{(-)}_{MKT}(n)$ are the partition functions under 
the $n$-fold iteration of the `MKT' decimation which is Tomboulis's 
modification of the MK RG.

If we assume for a moment that inequality (\ref{MKTineq}) holds {\it and} 
the MKT iteration leads eventually into the strong coupling regime, 
inequality (\ref{ineq}) follows, and this implies electric flux string 
formation and confinement in the sense that 't Hooft's string tension 
$\sigma_{tH}$ satisfies
\be
\sigma_{tH}>0 \quad \forall g^2\,,
\ee
where $g$ denotes the bare coupling constant.

One question that arises immediately is whether Ito's result, establishing 
flow to the strong coupling fixed point, also holds for Tomboulis's 
modification, which depends on an additional parameter $r=1-\epsilon$, 
$\epsilon>0$. We found that $r<1$ has the same effect as increasing the 
dimension and therefore for weak coupling the flow actually goes towards 
the {\it weak coupling fixed point}. 

\section{The fundamental issue}

As remarked before, the MK RG in 4$D$ shows {\it no structural difference}
between abelian (such as $U(1)$) and nonabelian (such as $SU(N)$) models:
the flow is always attracted by the strong coupling fixed point.
This was already pointed out in the seminal paper \cite{jose}, where
this insight was actually traced to Wilson's 1976 Carg\`ese lectures;
 as remarked, a proof of this fact was given by Ito \cite{ito}.  

This means that the original comparison argument given by Tomboulis 
{\it has to fail} for $U(1)$, because the $4D$ $U(1)$ model has vanishing 
string tension for sufficiently weak coupling. In fact, any similar 
argument that does not explicitly make use of the nonabelian nature of the 
gauge group has to fail. 

\section{Technical points}
{\it (a) The parameter $r$}.

The MKT decimation proceeds as follows: one starts with the character 
expansion of the plaquette coupling function (Gibbs factor)
\be
f(U)\equiv \exp A_p(U)=
F_0\left[1+\sum_{j\neq0}(2j+1)c_j(\beta)\chi_j(U)\right]\,.
\ee
In essence the decimation amounts to alternating raising the coupling 
function to the power $2^{D-2}$ and raising the Fourier coefficients
$c_j$ to the power $4r$. Explicitly
\be
f^{(n)}(U)\mapsto \frac{f^{(n)}(U)^4}{\int f^{(n)}(U)^4 dU}\equiv
g^{(n)}(U)\,,
\ee
\be
g^{(n)}(U)=1+\sum_{j>0} (2j+1)c_j(n) \chi_j(U)
\ee
\be
f^{(n+1)}(U)=1+\sum_{j>0} (2j+1)c_j(n)^{4r} \chi_j(U)\,.
\ee
Equality of the two exponents $2^{D-2}$ and $4r$ would mean that one is 
working in the critical dimension $D_c$ and one finds easily 
\be
D_c=4+\frac{\ln r}{\ln 2}<4\,,
\ee
so that with $r<1$ in $4D$ one is {\it above} the critical dimension and 
has to expect a phase transition. This is indeed the case; we have run 
the iteration for $r=0.9$ and two close values of $\beta\equiv 2/g^2$ and 
found a bifurcation of the flow: for $\beta=4.79$ the flow is attracted 
to the weak coupling fixed point, whereas for $\beta=4.80$ is flows to the 
strong coupling fixed point. The fact that for weak coupling the flow 
converges to the weak coupling fixed point can also be seen in a simple 
Gaussian approximation.

\begin{figure}[htb]
\includegraphics[width=.5\columnwidth]{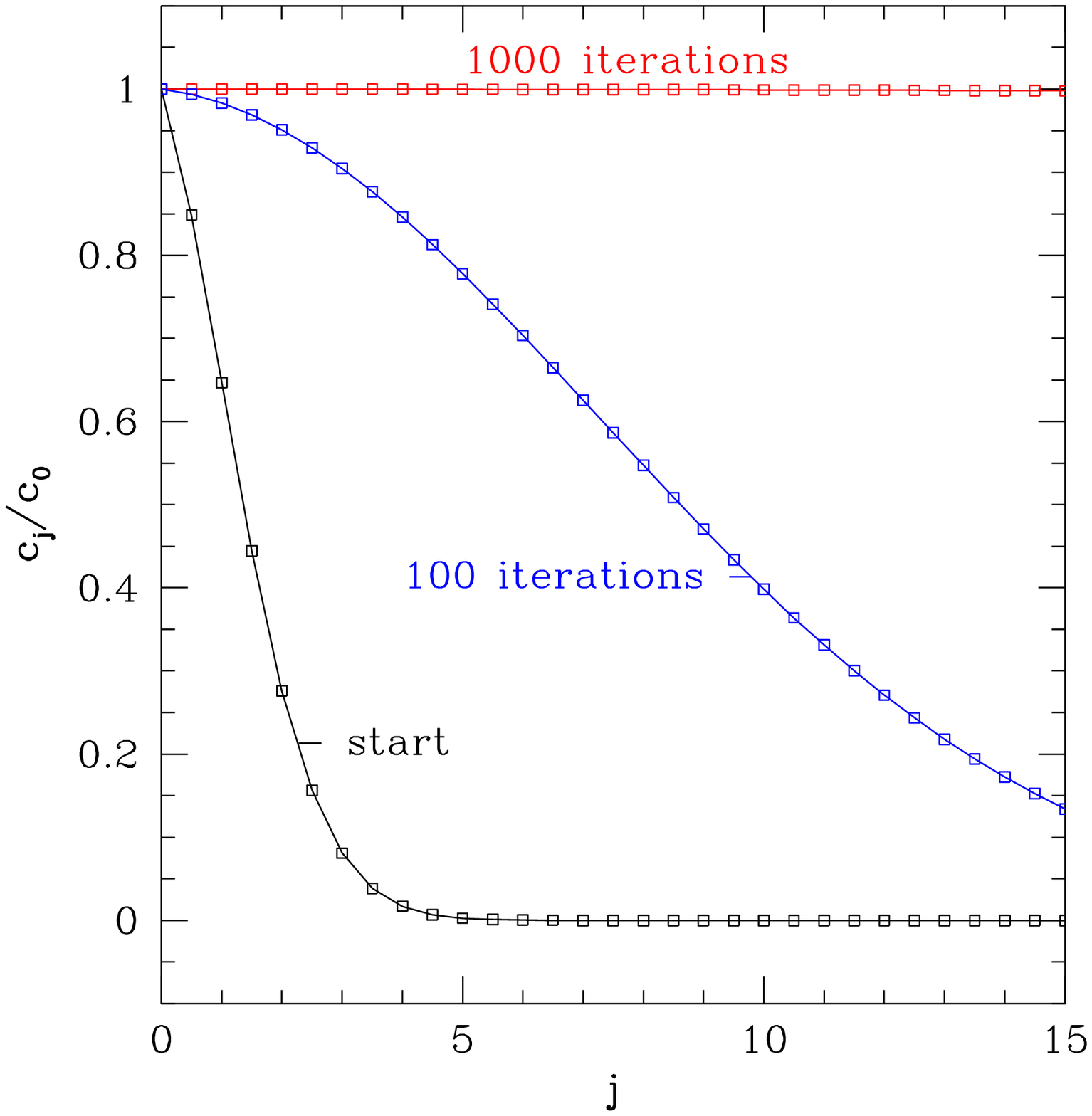}
\includegraphics[width=.5\columnwidth]{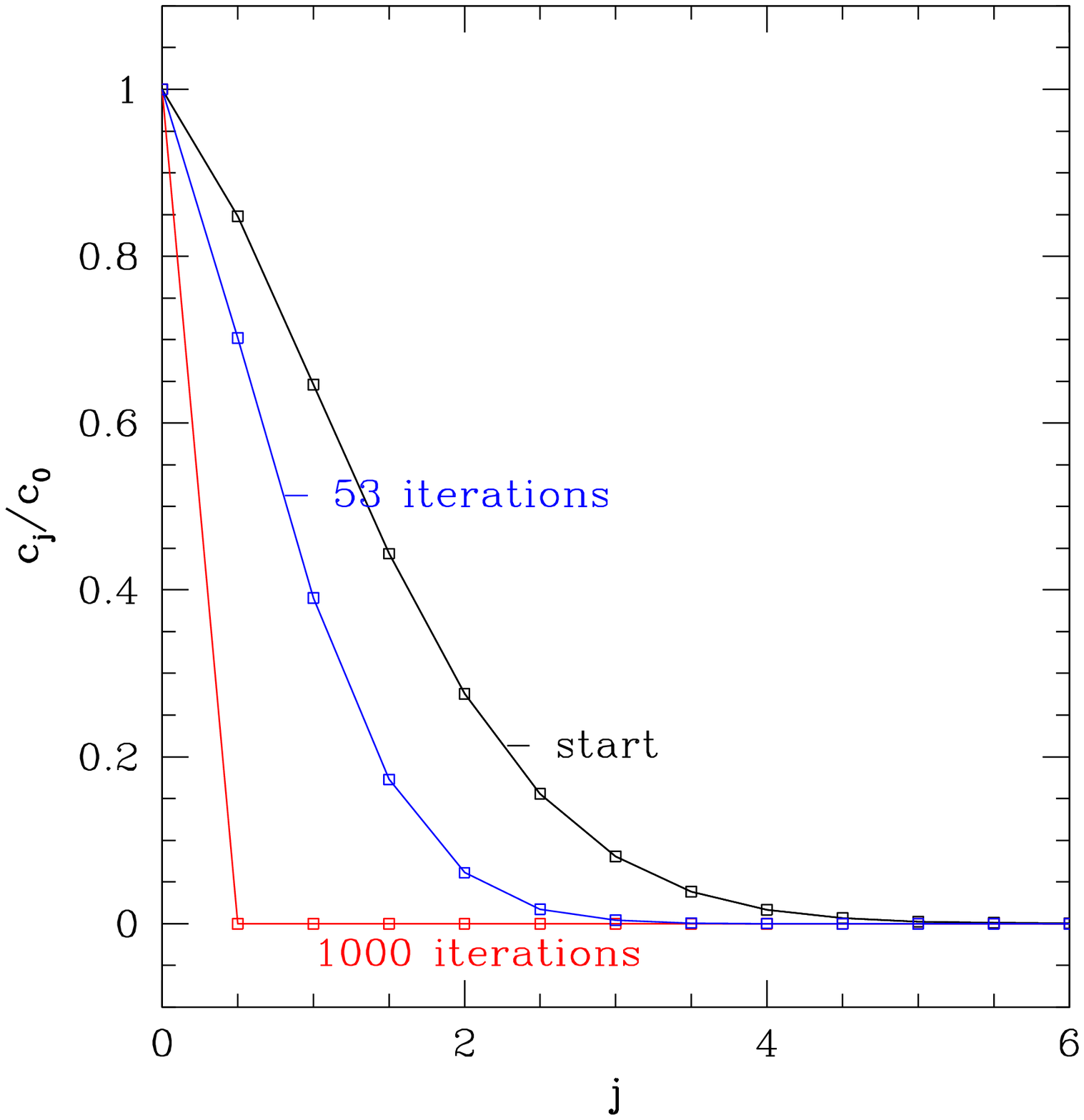}
\caption{Evolution of $c_j/c_0$ under Tomboulis' modified MK RG with
$r=0.9$. $\beta=4.80$ (left plot), $\beta=4.79$ (right plot); lines drawn
to guide the eye.
}
\label{plot}
\end{figure}

{\it (b) Existence of a common interpolation parameter $\alpha^\ast$ for 
$Z$ and $Z^{(-)}$.}

This is an essential point in Tomboulis's strategy. He has to find for all 
$n$ an $\alpha(n)^\ast\le 1-\delta$ such that 
\be
\frac{Z^{(-)}_\Lambda}{Z_\Lambda}=
\frac{Z_\Lambda^{(-)} (\{\alpha^*(n)c_{j}(n)\}}
{Z_\Lambda (\{\alpha^*(n)c_{j}(n)\})}
\ee
His argument (in Appendix C of \cite{tomb07}), based on the {\it implicit 
function theorem}, is flawed. He introduces a certain function 
$\Psi(\lambda,t)$ in terms of interpolated partition functions and 
is looking for a $t(\lambda)$ such that
\be
\Psi(\lambda,t)=0\,.
\ee
There is a solution $t_0$ at $\lambda=0$, but the sought after $\alpha^*$ 
would emerge from $t(1)$. Tomboulis is able to show that 
$\frac{\partial}{\partial t}\Psi(\lambda,t)\neq 0$, so by the implicit 
function theorem there is a solution near $\lambda=0$. But the information 
is not sufficient to allow the extension to $\lambda=1$, as shown 
by a simple counterexample due to T. Kanazawa \cite{kana}:
\be 
\Psi(\lambda,t)\equiv e^{-t}-1+2\lambda
\ee
which has the solution $t(\lambda)=-\log(1-2\lambda)$.

\section{Can the problems be fixed?} The choice of the parameter $r$ is 
very subtle, because one has to make sure of two things: (1) the 
decimation has to run into the strong coupling fixed point and (2) $r$ 
has to be kept away from 1, as is stressed in \cite{tomb07}.  This second 
issue is not discussed in \cite{tomb07} in a quantitative way, while the 
issue (1) is not addressed at all. Tomboulis hinted orally at the option 
of making $r$ dependent on $n$, the number of the iterations, but exactly 
how this would have to be done remains unclear.

In this respect the case of $U(1)$ is instructive: for $r=1$ the common 
interpolation parameter $\alpha^*$ cannot exist, because it would imply 
the existence of a nonvanishing string tension at all values of the bare 
coupling, in contradiction with proven facts (\cite{guth,fs}). 

Quite generally, we think that any strategy based on a Migdal-Kadanoff 
type decimation is very unlikely to succeed, because these hierarchical 
approximations do not show any structural difference between abelian (like 
$U(1)$) and nonabelian (like $SU(2)$) models.


\begin{thebibliography}{99}
\bibitem{yoneya} T.~Yoneya, Nucl. Phys. {\bf B144} (1978) 195.

\bibitem{thooft} G.~'t~Hooft, Nucl. Phys. {\bf B138} (1978) 1; 
          {\bf B153} (1979) 141.

\bibitem{mp} G.~Mack and V.~B.~Petkova, Ann. of Physics {\bf 123} (1980) 117.

\bibitem{wilson} K.~Wilson, Phys. Rev. {\bf D10} (1975)  2445.

\bibitem{polyakov} Phys.Lett.{\bf B 72} (1978) 477.

\bibitem{bs} C.~Borgs and E.~Seiler, Commun.Math.Phys. {\bf 91} (1983) 329.

\bibitem{ty} E.~T.~Tomboulis amd L.G. Yaffe, Commun.Math.Phys. 
{\bf 100} (1985,313.

\bibitem{tombprl} E.~T.~Tomboulis, Phys.Rev.Lett. {\bf 50} 
(1983) 885.

\bibitem {mk} A.~A.~Migdal, JETP {\bf 69} (1975) 810;
          L.~Kadanoff, Ann. Phys. {\bf 100} (1976) 359.

\bibitem{jose} J.~V.~Jose, L.~P.~Kadanoff, S.~Kirkpatrick, D.~E.~Nelson, 
Phys.Rev. {\bf B16} (1977) 1217.

\bibitem{ito} K.~R.~Ito, Phys.Rev.Lett.{\bf 54} (1985) 2383; 
Phys.Rev.Lett.{\bf 55} (1985) 558; 

\bibitem{guth} A.~Guth, Phys. Rev. D21, (1980) 2291. 

\bibitem{fs} J.~Fr\"ohlich and T.~Spencer, Commun.Math.Phys. {\bf 83} 
(1982) 411.

\bibitem{kana} T.~Kanazawa, private communication to K~.R.~Ito.

\bibitem{seiler}  E.Seiler, RIMS Kokyuroku (Res.Inst.Math.Sci., Kyoto 
    University) {\bf 1386} (2004), 193; arXiv hep-th/0312015.

\bibitem{itoconf} K.~R.~Ito, Lett.Math.Phys.{\bf 37} (1996) 349.

\bibitem{tomb07}  E.~T.~Tomboulis, Confinement for all values of the
  coupling in 4D  SU(2) gauge theory, arXiv:0707.2179[hep-th]; these 
  proceedings.

\bibitem{is} K.~R.~Ito and E.~Seiler, arXiv:0803.3019 [hep-th].

\end{thebibliography}
\end{document}